\documentclass{ws-ijmpa}

\begin{document}

\markboth{H.\,L.\,L.~Roberts, L.~Chang and C.\,D.~Roberts}
{Impact of DCSB on Meson Structure and Interactions}

%
\catchline{}{}{}{}{}
%

\title{IMPACT OF DYNAMICAL CHIRAL SYMMETRY BREAKING ON MESON STRUCTURE AND INTERACTIONS}

\author{H.\,L.\,L.~ROBERTS,$^{1,2}$ L.~CHANG\,$^{3}$ and C.\,D.~ROBERTS$^{1,2,4}$}

\address{$^1$Physics Division, Argonne National Laboratory\\ Argonne, Illinois 60439, USA\\
$^2$Institut f\"ur Kernphysik,
Forschungszentrum J\"ulich\\ D-52425 J\"ulich, Germany\\
$^3$Institute of Applied Physics and Computational Mathematics\\ Beijing 100094, China\\
$^4$Department of Physics, Peking University\\ Beijing 100871, China}

\maketitle

\begin{history}
\received{19 July 2010}
\end{history}

\begin{abstract}
We provide a glimpse of recent progress in meson physics made via QCD's Dyson-Schwinger equations with: a perspective on confinement and dynamical chiral symmetry breaking (DCSB); a pr\'ecis on the physics of in-hadron condensates; results for the masses of the $\pi$, $\sigma$, $\rho$, $a_1$ mesons and their first-radial excitations; and an illustration of the impact of DCSB on the pion form factor.
\end{abstract}

\ccode{PACS numbers:
12.38.Aw, 
12.38.Lg, 
14.40.Be, 
13.40.Gp. 
}

\vspace*{4ex}

\hspace*{-\parindent}\textbf{1.~Confinement, DCSB and in-hadron condensates}.
QCD is a theory whose elementary excitations are confined.  Moreover, less-than 2\% of a nucleon's mass can be attributed to the so-called current-quark masses that appear in QCD's Lagrangian, with the remainder being generated through dynamical chiral symmetry breaking (DCSB). Neither confinement nor DCSB is apparent in QCD's Lagrangian and yet they play the dominant role in determining the observable characteristics of real-world QCD.  The physics of hadrons is ruled by such \emph{emergent phenomena}, which can only be elucidated via nonperturbative methods in quantum field theory.  

\begin{figure}[t]
\includegraphics[clip,width=0.5\textwidth]{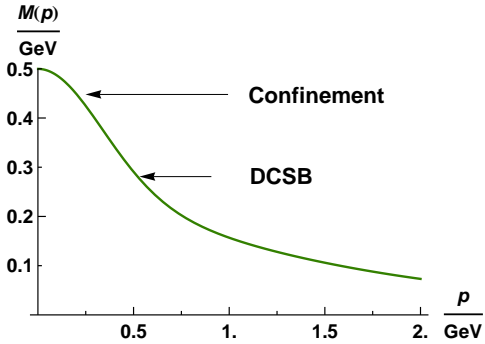}
\vspace*{-23ex}

\rightline{\parbox{0.45\textwidth}{
\caption{\label{MpFig} Dressed-quark mass function used, e.g., in an extensive study of nucleon electromagnetic form factors.\protect\cite{Cloet:2008re}  Dynamical chiral symmetry breaking is evident in the enhancement around $p=0.5\,$GeV; and confinement is signalled by the inflexion point at around $p=0.25$\,GeV.}}}\bigskip
\end{figure}

Confinement and DCSB can both be seen in the dressed-quark mass function, Fig.\,\ref{MpFig}, the behaviour of which is essentially nonperturbative for infrared momenta.  The existence of DCSB is signalled by the rapid increase in magnitude as $p$ decreases from $1 \to 0.5\,$GeV.  This prediction of Dyson-Schwinger equation (DSE) studies in QCD,\cite{Bhagwat:2003vw,Bhagwat:2006tu} is confirmed by simulations of lattice-QCD.\cite{Bowman:2005vx}  The reality of DCSB means that the Higgs mechanism is largely irrelevant to the normal matter in the universe.  

In analysing large momentum transfer processes it is common to use a light-front formulation of QCD.  This has merit but also, apparently, a serious drawback; viz., DCSB has not yet been realised on the light-front.  However, progress has recently been made,\cite{Brodsky:2010xf} by arguing for a shift in paradigm so that DCSB is understood as being expressed in properties of hadrons instead of the vacuum.  It is contended that: the impact of all condensates is entirely expressed in the properties of QCD's asymptotically realisable states; there is no evidence that vacuum condensates are anything more than mass-dimensioned parameters in one or another theoretical truncation scheme; and condensates do not describe measurable spacetime-independent configurations of QCD's elementary degrees-of-freedom in an hadron-less ground state.  This position assumes confinement, from which it follows that all observable consequences of QCD can, in principle, be computed using an hadronic basis.

\begin{figure}[t]
\includegraphics[clip,width=0.43\textwidth]{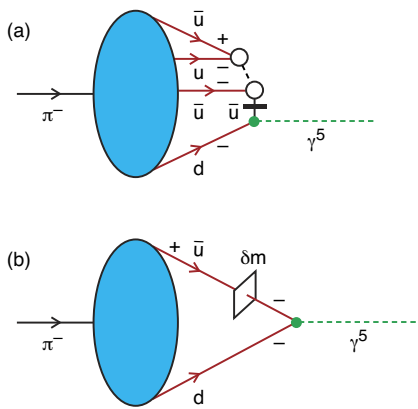}
\vspace*{-33ex}

\rightline{\parbox{0.45\textwidth}{\caption{\label{instantaneous}
Light-front contributions to $\rho_\pi=-\langle 0| \bar q \gamma_5 q |\pi\rangle$.
\emph{Upper panel} -- A non-valence piece of the meson's light-front wavefunction, whose contribution to $\rho_\pi$ is mediated by the light-front instantaneous quark propagator (vertical crossed-line).  The ``$\pm$'' denote parton helicity.
\emph{Lower panel} -- There are infinitely many such diagrams, which can introduce chiral symmetry breaking in the light-front wavefunction in the absence of a current-quark mass.
(The case of $f_\pi$ is analogous.)}}}\vspace*{4ex}
\end{figure}

This suggests a solution to the problem of DCSB in the light-front formulation, illustrated in Fig.\,\ref{instantaneous}.  The light-front-instantaneous quark propagator can mediate a contribution from higher Fock states to the matrix elements that define the pion's pseudoscalar ($\rho_\pi$) and pseudovector ($f_\pi$) decay constants.  Such diagrams connect DCSB-components of the meson's light-front wavefunction to these matrix elements.  There are infinitely many such contributions and they do not depend sensitively on the current-quark mass in the neighborhood of the chiral limit.  This leads to a conjecture that DCSB on the light-front is expressed via in-hadron condensates and is connected with sea-quarks derived from higher Fock states.

Confinement is signalled in Fig.\,\ref{MpFig} by the inflexion point in $M(p)$ at $p\approx 0.25$\,GeV.  This relationship is explained elsewhere.\cite{Krein:1990sf,Roberts:2007ji,Chang:2010jq}  Confinement and DCSB are intimately connected.  It is natural to ask whether the connection is accidental or causal.  There are models with DCSB but not confinement, however, a model with confinement but lacking DCSB has not yet been identified (see Secs.\,2.1 and 2.2 of Ref.\,[\refcite{Roberts:2007jh}]).  This leads to a conjecture that DCSB is a necessary consequence of confinement.  

These observations highlight that confinement is related to the analytic properties of QCD's Schwinger functions.\cite{Krein:1990sf,Roberts:2007ji,Chang:2010jq}  The question of light-quark confinement can thus be translated into the challenge of charting the infrared behavior of QCD's \emph{universal} $\beta$-function.  This well-posed problem can be addressed in any framework enabling the nonperturbative evaluation of renormalisation constants.

\smallskip

\hspace*{-\parindent}\textbf{2.~DCSB and the hadron spectrum}.
Through the gap and Bethe-Salpeter equations (BSEs) the pointwise behaviour of the $\beta$-function determines the pattern of chiral symmetry breaking; e.g., the behaviour in Fig.\,\ref{MpFig}.  Moreover, the fact that the DSEs connect the $\beta$-function to experimental observables entails, e.g., that comparison between computations and observations of the hadron mass spectrum, and hadron elastic and transition form factors, can be used to chart the $\beta$-function's long-range behaviour.  In order to realise this goal, a nonperturbative symmetry-preserving DSE truncation is necessary.  Steady quantitative progress can be made with a scheme that is systematically improvable \cite{Bender:1996bb}.  In fact, the mere existence of such a scheme has enabled the proof of exact nonperturbative results in QCD.\cite{Holl:2005vu,Bhagwat:2007ha}

Alternatively, significant advances in understanding the essence of QCD could be made with symmetry-preserving kernel \emph{Ans\"atze} that express important additional nonperturbative effects, which are impossible to capture in any finite sum of contributions. Such an approach is now available\cite{Chang:2009zb}.  It begins with a novel form for the axial-vector BSE, which is valid when the quark-gluon vertex is fully dressed.  Therefrom, a Ward-Takahashi identity for the Bethe-Salpeter kernel is derived and solved for a class of dressed-quark-gluon-vertices.  The solution yields a symmetry-preserving closed system of gap and vertex equations.  The analysis can readily be extended to the vector BSE, so a comparison is possible between the responses of parity-partners in the meson spectrum to nonperturbatively dressing the quark-gluon vertex.  The results indicate that: spin-orbit splitting between ground-state mesons is dramatically enhanced by DCSB;\cite{Chang:2009zb} DCSB generates a large anomalous chromomagnetic moment as an essential part of the dressed-quark-gluon vertex;\cite{Chang:2010jq} and, owing to the symmetry-preserving nature of the truncation procedure, the $M(p)$-driven vertex corrections alter the Bethe-Salpeter kernel in such a way as to leave ground-state pseudoscalar and vector meson masses almost unchanged.\cite{Chang:2010jq}

Motivated by this ongoing work, we employed the DSE model formulated in Ref.\,[\refcite{GutierrezGuerrero:2010md}] to compute the masses of some meson ground-states and their radial excitations, with the results presented in Table\,\ref{masses}.  The first row lists masses obtained in rainbow-ladder truncation.  Of course, given the symmetry preserving nature of the truncation, in the chiral-limit $m_\pi=0$ and $m_\sigma = 2 M$ (which follows from Eq.\,(14) in Ref.\,[\refcite{GutierrezGuerrero:2010md}]), where $M=0.40\,$GeV is chiral-limit value of the model's momentum-independent dressed-quark mass.  The second row lists values obtained with spin-orbit repulsion added to the scalar and axial-vector channels, the strength of which is described by a single parameter, $g_{\rm SO}=0.29$, fixed so as to yield $m_{a_1/f_1}=1.243\,$GeV.  (We emphasise that introducing $g_{\rm SO}\neq 1$ is an expedient that mimics those effects of a \emph{momentum-dependent} mass function exposed and quantified in Refs.\,[\refcite{Chang:2010jq,Chang:2009zb}].)  It will be noted that $m_\sigma$ increases to a value which matches an estimate for the $\bar q q$-component of this meson obtained using unitarised chiral perturbation theory.\cite{Pelaez:2006nj}

\begin{table}[t]
\tbl{\label{masses}
%
Meson masses (GeV) computed using a contact-interaction DSE kernel,\protect\cite{GutierrezGuerrero:2010md} which produces a momentum-independent dressed-quark mass $M=0.41\,$GeV from a current-quark mass of $m=8\,$MeV.  ``RL'' denotes rainbow-ladder truncation.\protect\cite{Bender:1996bb}  Row-2 is obtained by augmenting the RL kernel with spin-orbit repulsion (see text); and Row-3 lists experimental masses\protect\cite{Amsler:2008zzb} for comparison. NB.\ We implement isospin symmetry so, e.g., $m_\omega=m_\rho$, $m_{f_1}=m_{a_1}$, etc.}
{\hspace*{2em}\begin{tabular}{@{}lccccccccc@{}}\toprule
   & $m_\pi$ & $m_\rho$ & $m_\sigma$ & $m_{a_1}$
   & $m_{\pi^\ast}$ & $m_{\rho^\ast}$ & $m_{\sigma^\ast}$ & $m_{a_1^\ast}$ \\
RL & 0.141 & 0.798 & 0.825 & 1.073 & 1.434 & 1.434 & 1.479 & 1.461 \\
RL $+ \,g_{\rm SO}$ & 0.141 & 0.798 & 1.079 & 1.243
                    & 1.434 & 1.434 & 1.487 & 1.485\\\hline
experiment & 0.140 & 0.777 & 0.4 - 1.2 &  1.243 &1.3$\,\pm\,$0.1 & 1.465 & 0.980 & 1.426
\end{tabular}\hspace*{2em}}
\end{table}

It is worth emphasising that in quantum field theory as in quantum mechanics, the bound-state amplitude for a first radial excitation must possess a single zero.\cite{Holl:2005vu}  We implement that feature via the method of Ref.\,[\refcite{Volkov:1999yi}]; so that, e.g., the BSE for the $\rho$-meson's first radial excitation is the following eigenvalue problem for $P^2=-m_{\rho^\ast}^2$:
\begin{equation}
1 = \frac{1}{3\pi^2m_G^2} \int_0^1 \!\! d\alpha\; 2\!\int_q^\Lambda
\frac{\frac{1}{2}q^2 + M^2 - \alpha (1-\alpha) P^2}{(q^2 + M^2 + \alpha (1-\alpha) P^2)^2} \, (1 - d_{\rm r} q^2)\,,
\end{equation}
with:\cite{GutierrezGuerrero:2010md} $m_G=0.11\,$GeV, a gluon mass-scale; $M$ the model's computed dressed-quark mass; $\int_q^\Lambda := \int d^4 q/(4\pi^2)$, regularised as explained therein; and $d_{\rm r} = 1/[2 \Lambda_{\rm ir}^2]$, where $1/\Lambda_{\rm ir} = 0.83\,$fm is the model's confinement length-scale.  Our results indicate that the impact of $M(p^2)$-driven vertex corrections can be large on ground-state masses but is much diminished for meson radial excitations.

\smallskip

\begin{figure}[t] 
\includegraphics[clip,width=0.40\textwidth,angle=-90]{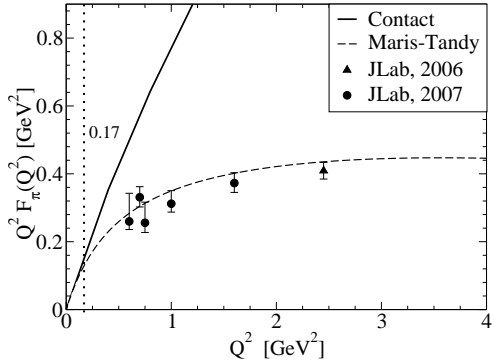}
\vspace*{-31.5ex}

\rightline{\parbox{0.45\textwidth}{
\caption{\label{figFpi} \emph{Solid curve}: $Q^2 F_{\pi}(Q^2)$ obtained
using a contact interaction.\protect\cite{GutierrezGuerrero:2010md}  \emph{Dashed curve}: DSE prediction \protect\cite{Maris:2000sk}, which employed a momentum-de\-pen\-dent renormalisation-group-im\-pro\-ved gluon exchange interaction. For $Q^2>0.17\,$GeV$^2\approx M^2$, marked by the vertical \emph{dotted line}, the contact interaction result for $F_{\pi}(Q^2)$ differs from that in Ref.\,[\protect\refcite{Maris:2000sk}] by more than 20\%.  The data are from Refs.\,[\protect\refcite{Horn:2006tm,Tadevosyan:2007yd}].}}}\vspace*{4ex}
\end{figure}

\hspace*{-\parindent}\textbf{3.~Elastic form factors}.
Once one has bound-state masses and Bethe-Salpeter amplitudes, the computation of meson elastic and transition form factors can proceed.  DSE calculations\cite{Maris:1998hc,Maris:2000sk} of $F^{\rm em}_\pi(Q^2)$ are an archetype ($Q^2$ is the squared-momentum-transfer).  The most systematic\cite{Maris:2000sk} predicted the measured form factor.\cite{Volmer:2000ek}  An elucidation of the sensitivity of $F^{\rm em}_\pi$ to the pointwise behaviour of the interaction between quarks is the main theme of Ref.\,[\refcite{GutierrezGuerrero:2010md}].  In Fig.\,\ref{figFpi} we compare $F^{\rm em}_\pi$ computed using a contact-interaction with the QCD-based DSE prediction\cite{Maris:2000sk} and contemporary experimental data.\cite{Volmer:2000ek,Horn:2006tm,Tadevosyan:2007yd}  Both the QCD-based result and that obtained from the momentum-in\-de\-pen\-dent interaction yield the same values for the pion's static properties. However, for $Q^2>0$ the form factor computed using $\sim 1/k^2$ vector boson exchange is immediately distinguishable from that produced by a momentum-independent interaction.  The figure shows that for $F_\pi^{\rm em}$, existing experiments can already distinguish between different possibilities for the quark-quark interaction.

The analysis of Ref.\,[\refcite{GutierrezGuerrero:2010md}] demonstrates that when a momentum-independent vector-exchange interaction is regularised in a symmetry-preserving manner, the results are directly comparable with experiment, computations based on well-defined and systematically-improvable truncations of QCD's DSEs\cite{Maris:2000sk}, and perturbative QCD.  However, a contact interaction, whilst capable of describing pion static properties well, generates a form factor whose evolution with $Q^2$ deviates markedly from experiment for $Q^2>0.17\,$GeV$^2\approx M^2$ and produces asymptotic power-law behaviour: $Q^2 F_\pi(Q^2)\approx\,$constant, in serious conflict with perturbative-QCD \cite{Farrar:1979aw,Efremov:1979qk}.

Notably, the contact interaction produces a momentum-independent dressed-quark mass function, in contrast to QCD-based DSE studies\cite{Bhagwat:2003vw,Bhagwat:2006tu,Roberts:2007ji} and lattice-QCD.\cite{Bowman:2005vx}  This is the origin of the discrepancy between the form factor it produces and extant experiment.  Hence Ref.\,[\refcite{GutierrezGuerrero:2010md}] highlights that form factors are capable of mapping the running of the dressed-quark mass function.  Efforts are underway to establish the signals of the running mass in baryon elastic and transition form factors.  In this connection one statement can readily be made.  Bethe-Salpeter amplitudes are peaked at zero relative momentum.  Hence, the domain of greatest support in the impulse approximation calculation of meson elastic form factors is that with each quark absorbing $Q/2$.\cite{Roberts:1994hh}  The dressed-quark mass, $M(p^2)$, is perturbative for $p^2>2\,$GeV$^2$.  It can be argued from these observations that pQCD behaviour will not be observed in meson form factors unless $Q^2>8\,$GeV$^2$.

\smallskip

\hspace*{-\parindent}\textbf{4.~Epilogue}.
We have exemplified the impact of DCSB upon observables.  The behaviour of the $M(p)$ heralds DCSB; and the $p$-dependence manifest in Fig.\,\ref{MpFig} is an essentially quantum field theoretical effect.  Elucidating its consequences therefore requires a nonperturbative, symmetry-preserving approach, where the latter means preserving Poincar\'e covariance, chiral and electromagnetic current-conservation, etc.  The DSEs provide such a framework.  Experimental and theoretical studies are underway that will identify observable signals of $M(p)$ and thereby explain the most important mass-generating mechanism for visible matter in the Standard Model.

This is an exciting time.  Through the DSEs, we are positioned to unify phenomena as apparently diverse as the: hadron spectrum; hadron elastic and transition form factors, from small- to large-$Q^2$; and parton distribution functions.\cite{Holt:2010vj}  The key is an understanding of both the fundamental origin of nuclear mass and the far-reaching consequences of the mechanism responsible; namely, DCSB.  These things might lead us to an explanation of confinement, the phenomenon that makes nonperturbative QCD the most interesting piece of the Standard Model.

\smallskip

\hspace*{-\parindent}\textbf{Acknowledgments}.
We acknowledge valuable discussions with C.~Hanhart, S.\,M.~Schmidt and P.\,C.~Tandy.
This work was supported by:
Forschungszentrum J\"ulich GmbH (HLLR, CDR);
the National Natural Science Foundation of China, contract no.\ 10705002 (LC);
the U.\,S.\ Department of Energy, Office of Nuclear Physics, contract no.~DE-AC02-06CH11357 (HLLR, CDR);
and the Department of Energy's Science Undergraduate Laboratory Internship programme (HLLR).
%


\begin{thebibliography}{00}


\bibitem{Cloet:2008re}
I.~C.~Clo{\"{e}}t \emph{et al}., 
  Roberts, {\em Few Body Syst.} {\bf 46}, 1 (2009).

\bibitem{Bhagwat:2003vw}
M.~S. Bhagwat, M.~A. Pichowsky, C.~D. Roberts and P.~C. Tandy, {\em Phys. Rev.}
  {\bf C68}, 015203 (2003).

\bibitem{Bhagwat:2006tu}
M.~S. Bhagwat and P.~C. Tandy, {\em AIP Conf. Proc.} {\bf 842}, 225 (2006).

\bibitem{Bowman:2005vx}
P.~O. Bowman {\em et~al.}, {\em Phys. Rev.} {\bf D71}, 054507 (2005).

\bibitem{Krein:1990sf}
  G.~Krein, C.~D.~Roberts and A.~G.~Williams,
  \emph{Int.\ J.\ Mod.\ Phys}.\  {\bf A7}, 5607 (1992).

\bibitem{Roberts:2007ji}
C.~D. Roberts, {\em Prog. Part. Nucl. Phys.} {\bf 61}, 50 (2008).

\bibitem{Chang:2010jq}
  L.~Chang and C.~D.~Roberts,
  ``Hadron Physics: The Essence of Matter,''
  arXiv:1003.5006 [nucl-th].

\bibitem{Roberts:2007jh}
  C.~D.~Roberts, M.~S.~Bhagwat, A.~H{\"{o}}ll and S.~V.~Wright,
  \emph{Eur.\ Phys.\ J.\ ST} {\bf 140}, 53 (2007).

\bibitem{Brodsky:2010xf}
  S.~J.~Brodsky, C.~D.~Roberts, R.~Shrock and P.~C.~Tandy,
  ``Essence of the vacuum quark condensate,''
  arXiv:1005.4610 [nucl-th].

\bibitem{Bender:1996bb}
  A.~Bender, C.~D.~Roberts and L.~Von Smekal,
  \textit{Phys.\ Lett}.\  {\bf B380}, 7 (1996).

\bibitem{Holl:2005vu}
  A.~H{\"o}ll \emph{et al}., 
  \emph{Phys.\ Rev}.\  {\bf C71}, 065204 (2005).

\bibitem{Bhagwat:2007ha}
  M.~S.~Bhagwat \emph{et al}., 
  \emph{Phys.\ Rev}.\  C {\bf 76}, 045203 (2007).

\bibitem{Chang:2009zb}
L.~Chang and C.~D. Roberts, {\em Phys. Rev. Lett.} {\bf 103}, p. 081601 (2009).

\bibitem{GutierrezGuerrero:2010md}
  L.~X.~Guti{\'e}rrez-Guerrero, A.~Bashir, I.~C.~Clo{\"{e}}t and C.~D.~Roberts,
  \emph{Phys.\ Rev}.\  {\bf C81}, 065202 (2010).

\bibitem{Amsler:2008zzb}
  C.~Amsler {\it et al.}  [Particle Data Group],
  \emph{Phys.\ Lett}.\  B {\bf 667}, 1 (2008).

\bibitem{Pelaez:2006nj}
  J.~R.~Pel{\'{a}}ez and G.~R{\'{\i}}os,
  \emph{Phys.\ Rev.\ Lett}.\  {\bf 97}, 242002 (2006)

\bibitem{Volkov:1999yi}
  M.~K.~Volkov and V.~L.~Yudichev,
  \emph{Phys.\ Part.\ Nucl}.\  {\bf 31}, 282 (2000)
  [\emph{Fiz.\ Elem.\ Chast.\ Atom.\ Yadra} {\bf 31}, 576 (2000)].

\bibitem{Maris:1998hc}
  P.~Maris and C.~D.~Roberts,
  \emph{Phys.\ Rev}.\  {\bf C58}, 3659 (1998).

\bibitem{Maris:2000sk}
  P.~Maris and P.~C.~Tandy,
  \emph{Phys.\ Rev}.\  {\bf C62}, 055204 (2000).

\bibitem{Volmer:2000ek}
  J.~Volmer {\it et al.},  
  \emph{Phys.\ Rev.\ Lett}.\  {\bf 86}, 1713 (2001).

\bibitem{Horn:2006tm}
  T.~Horn {\it et al.},  
  \emph{Phys.\ Rev.\ Lett}.\  {\bf 97}, 192001 (2006).

\bibitem{Tadevosyan:2007yd}
  V.~Tadevosyan {\it et al.},  
  \emph{Phys.\ Rev}.\  {\bf C75}, 055205 (2007).

\bibitem{Farrar:1979aw}
  G.~R.~Farrar and D.~R.~Jackson,
  \emph{Phys.\ Rev.\ Lett}.\  {\bf 43}, 246 (1979).

\bibitem{Efremov:1979qk}
  A.~V.~Efremov and A.~V.~Radyushkin,
  \emph{Phys.\ Lett}.\  {\bf B94}, 245 (1980).

\bibitem{Roberts:1994hh}
  C.~D.~Roberts,
  \emph{Nucl.\ Phys}.\  A {\bf 605}, 475 (1996).

\bibitem{Holt:2010vj}
  R.~J.~Holt and C.~D.~Roberts,
  ``Distribution Functions of the Nucleon and Pion in the Valence Region,''
  arXiv:1002.4666 [nucl-th].

\end{thebibliography}
\end{document}